\documentstyle[sprocl]{article}

\input{epsf}

\def\be{\begin{equation}}
\def\ee{\end{equation}}
\def\bea{\begin{eqnarray}}
\def\eea{\end{eqnarray}}

\def\pp{
\setbox0=\hbox{$\frac{d\omega}{\pi}$}         
\dimen0=\wd0                                  
\rlap{\hbox to \dimen0{\hfil\hglue 0.7mm $-$\hfil}}
\int }                                        

\begin{document}

\title{SELF-CONSISTENT APPROXIMATIONS:  APPLICATION TO A QUASIPARTICLE
DESCRIPTION OF THE THERMODYNAMIC PROPERTIES OF RELATIVISTIC PLASMAS}

\author{Beno\^\i t Vanderheyden}

\address{The Niels Bohr Institute, Blegdamsvej 17, DK-2100 Copenhagen
N, Denmark}

\author{Gordon Baym}

\address{Department of Physics, University of Illinois at Urbana-Champaign,\\
1110 W. Green St., Urbana, Illinois 61801, USA}


\maketitle\abstracts{We generalize the concept of conserving,
  $\Phi$-derivable, approximations to relativistic field
  theories. Treating the interaction field as a dynamical degree of
  freedom, we derive the thermodynamical potential in terms of fully
  dressed propagators, an approach which allows us to resolve the
  entropy of a relativistic plasma into contributions from its
  interacting elementary excitations.  We illustrate the derivation
  for a hot QED plasma of massless particles.  We also discuss how the
  self-consistency of the treatment manifests itself into
  relationships between the contributions from interaction and matter
  fields.  }

\section{INTRODUCTION}

    Nuclear matter at very high densities is expected, because QCD is
asymptotically free, to be in the form of a plasma of deconfined quarks and
gluons.~\cite{CollPer75} Such plasmas were present in the early universe and
may exist in the cores of neutrons stars; current experiments aim to produce
and study them in collisions of ultrarelativistic nuclei.~\cite{QM96} In this
talk, we address the question of providing a microscopic description of the
thermodynamic properties of relativistic plasmas.  The elementary excitations
of weakly coupled quark-gluon plasmas are in fact, up to color factors, very
similar in character to those of relativistic electromagnetic plasmas.  Here,
for simplicity we confine ourselves to the thermodynamic properties of QED
plasmas.  Unless otherwise stated, the properties of the electron and photon
degrees of freedom which we consider can easily be translated into those of
quark and gluon degrees of freedom in QCD plasmas.

    Relativistic plasmas exhibit many of the familiar many-body effects
encountered in non-relativistic condensed matter systems, including screening
and the existence of collective modes.~\cite{reviewcollective} Debye screening
shields the static electric component of the interaction but does not affect
static magnetic fields.  Screening of finite frequency fields arises from
Landau damping mechanisms in a manner similar to the anomalous skin effect in
metals.~\cite{reuter} Early studies of relativistic plasmas have demonstrated
the existence of bosonic and fermionic collective
modes.~\cite{reviewcollective} The bosonic excitations are the familiar
longitudinal and transverse plasmon oscillations and appear as poles in the
photon propagator.  The fermionic collective excitations are unique to
relativistic systems and develop as poles in the low momentum region of the
electron propagator.  Investigations of the fermionic spectrum at one-loop
order have revealed the existence of a rich structure, including a gap at zero
momentum and two distinct excitation branches at small
momenta.~\cite{fermions}

    In the limit of small coupling constant $g$, these medium effects enter
the theory through one-loop order corrections, which are systematically
implemented by the ``Hard-Thermal-Loop'' (HTL) perturbation scheme of Braaten
and Pisarski.~\cite{HTL} In a plasma of massless particles at temperature $T$,
medium effects generically develop over the long wavelength scale $\sim 1/(g
T)$.  (By comparison, the interparticle spacing is $\sim 1/T$.)  Difficulties,
whose solution lies beyond the HTL scheme, arise from the lack of static
screening of magnetic interactions.  The long-ranged nature of static magnetic
fields causes the fermion quasiparticle damping rate to be logarithmically
divergent at finite temperature in perturbation theory.  Blaizot and Iancu
have shown that in high temperature QED inclusion of multiple scattering {\it
\`a la} Bloch-Nordsieck leads to well-defined, divergence-free quasiparticle
modes.~\cite{irfermion} In high temperature QCD, magnetic screening may
actually appear at the scale $1/g^2 T$ due to gluon self-interactions.  The
physics of the strong interactions at this scale is however non-perturbative
and is thus again beyond the reach of HTL scheme.  Related infrared
difficulties due to long-ranged gauge fields arise in current issues of
condensed matter theory, for instance in gauge field models of high $T_c$
superconductors and of the fractional quantum Hall
effect.~\cite{HalpLeeRead95}

    In this talk, we present a general framework for analyzing the effects of
the gap in the fermion spectrum and of the long-ranged gauge fields on the
thermodynamic properties of relativistic plasmas.  Since we are dealing with
interacting degrees of freedom, care is needed to avoid overcounting.  We
generalize to relativistic systems the $\Phi$-derivable conserving
approximations developed and Baym~\cite{Phiconser2} (see also
\cite{Phiconser1}) in the context of quantum transport theories.  As we shall
see, this formulation permits the resolution of the entropy of a relativistic
plasma into components relative to its elementary excitations.  We will also
show that the self-consistency of the approach implies subtle relationships
between the contributions to the entropy from matter and interaction fields;
these relationships are the manifestation of a proper counting of the degrees
of freedom.

\section{FROM THE THERMODYNAMICAL POTENTIAL TO THE ENTROPY}

    We illustrate the derivation for a hot relativistic QED plasma, a gas of
electrons and positrons in equilibrium with photons at temperature $T \gg m$,
the electron mass.  For simplicity, we set $m=0$.  At finite temperature, the
system is not Lorentz invariant; only rotational invariance is obeyed.  There
is thus a preferred frame -- the rest frame of the system -- in which medium
effects act separately on the longitudinal and transverse components of the
interaction.  It is thus natural to choose the Coulomb gauge.  We work in the
imaginary time formalism.

    As we have shown~\cite{VanBaym} the free energy $\Omega$ can be written in
terms of the fully dressed electron and photon propagators $G$ and $D$ as
\begin{eqnarray}
\Omega/ T&=&\Phi[G,D]-{\rm Tr\,} \Sigma G+
{\rm Tr\,} \log(-\gamma^0 G)+{1\over 2} {\rm Tr\,}\Pi D
-{1\over 2}{\rm Tr\,}\log(-D).
\label{W:QED}
\end{eqnarray}
The traces are over the four momenta and over spin and polarization
states, so for instance
\begin{eqnarray}
{\rm Tr\,}{\Sigma G} \equiv {\rm tr}\sum_{p,n} \Sigma(\omega_n,p)
G(\omega_n,p), \quad
{\rm Tr\,}{\Pi D} \equiv {\rm tr}\sum_{q,n} \Pi(\omega_n,q) D(\omega_n,q),
\end{eqnarray}
where $\omega_n=(2 n+1) i \pi T$ and $\omega_n=2 n i \pi T$ are the
electron and photon Matsubara frequencies, and ${\rm tr}$ represents
the traces over spins or polarizations. The second and third terms in
Eq.~(\ref{W:QED}) represent a contribution from the electron degrees
of freedom and have a form similar to those encountered in
non-relativistic version of conserving
approximations.~\cite{Phiconser2}  The generalization to relativistic
systems is implemented here by the last two terms, which correspond to
treating the electromagnetic interaction as a dynamical degree of
freedom.  The coupling between electron and photon modes is described
by the functional $\Phi[G,D]$, the sum of all
two-particle-irreducible skeleton diagrams, expressed in terms of
fully dressed $G$ and $D$ instead of the bare electron and photon
propagators $G_0$ and $D_0$.  Under a simultaneous variation of the
propagators $G$ and $D$,
\begin{eqnarray}
\delta \Phi[G,D]&=&{\rm Tr\,} \Sigma \delta G-{1\over 2} {\rm Tr\,}\Pi
\delta D,
\label{deltaphi:QED}
\end{eqnarray}
as one easily sees by removing an electron or photon line in a
given diagram contributing to $\Phi$.  The electron self-energy
$\Sigma[G,D]$ and the polarization operator $\Pi[G,D]$ are here
functionals of $G$ and $D$ and satisfy Dyson's equations
\begin{eqnarray}
G^{-1} = G_0^{-1}-\Sigma, \quad
D^{-1} = D_0^{-1}-\Pi.
\label{Dyson}
\end{eqnarray}

We emphasize that the representation of the thermodynamical potential in
Eq.~(\ref{W:QED}) is exact. This result is based upon two essential
properties of the $\Phi$ functional. The first given in
Eq.~(\ref{deltaphi:QED}), implies that the potential $\Omega$ is
stationary under variations of $G$ and $D$ that do not modify the free
propagators $G_0$ and $D_0$. The second property follows from the topology of
the diagrams contributing to $\Phi$. As each vertex of a given $\Phi$-diagram
is connected to two electron lines and one photon line, the functional $\Phi$
remains constant under the scaling transformations
\begin{eqnarray}
\Phi[G,D;g]&=&\Phi[s^{-f} G,s^{ 2f-2 v}D;s^v g], \label{scale2:QED}
\end{eqnarray}
where $g$ is the coupling constant at a vertex. To gain insight into
the meaning of this property, we note that when combined with
Eq.~(\ref{deltaphi:QED}) it implies ${\rm Tr\,} \Sigma G=- {\rm Tr\,}
\Pi D$.  This is one example of the interdependencies that arise
between the electron and photon contributions to thermodynamic
quantities when treated self-consistently.  The traces ${\rm Tr\,}
\Sigma G$ and $- {\rm Tr\,} \Pi D$ both represent the interaction
energy between electrons and photons. These expressions are equal
if $\Phi$ satisfies the scaling invariance of Eq.~(\ref{scale2:QED}).

The principle behind the conserving approximations consists of choosing a
physically motivated subset of diagrams contributing to $\Phi$, and deriving
the corresponding self-energy functionals via Eq.~(\ref{deltaphi:QED}).
Dyson's equations then provide self-consistent equations for the electron and
photon propagators. These approximations preserve the functional and
topological properties of $\Phi$, and a proper counting of the degrees of
freedom is ensured.  In particular, it follows from the stationarity
property of the thermodynamical potential $\Omega$ that the approximations
obtained for the propagators $G$ and $D$ lead to current densities and an
energy momentum tensor that obey the continuity equations for the
conservation of charge, momentum, and energy~\cite{Phiconser2}.

    As we shall see shortly, the entropy $S= -\partial\Omega/\partial
T|_{\mu,V}$ is a quantity well-suited to be analyzed in terms of its
quasiparticle content.  Converting the frequency sums into integrals in the
usual way\footnote{Due to the zero Matsubara mode, the sum over photon
frequencies becomes a principal value integral.}, we write $S$ as
\begin{eqnarray}
S=-\left.{\partial\Omega\over\partial T}\right|_{\mu,V}&=&S_f+S_b+S',
\label{ssfsbs}
\end{eqnarray}
where
\begin{eqnarray}
S_f&\equiv&-\sum_{p\pm}\int_{-\infty}^{\infty}{d\omega_p\over  \pi}
\,{\partial f\over \partial
T}\left( {\rm Im} \Sigma_\pm\, {\rm Re} G_\pm+{\rm Im}\,
\log(-G_\pm^{-1})\right),\label{sf} \\
S_b&\equiv&-\sum_{{q},l}\pp_{0}^{\infty}{d\omega_q\over \pi} \,{\partial
n\over \partial T} \left({\rm Im}\Pi_l\,{\rm Re} D_l+{\rm Im}
\log(-D_l^{-1})\right)
,\label{sb} \\
S'&\equiv&-{\partial (T \Phi)\over\partial T}
-\sum_{p\pm}\int_{-\infty}^\infty{d\omega_p\over  \pi}
\,{\partial f\over \partial T}  {\rm Im} G_\pm \,{\rm Re} \Sigma_\pm
 -\sum_{{q},l}  \pp_0^\infty{d\omega_q\over \pi}\, {\partial n\over \partial
 T}\,{\rm Im} D_l \,{\rm Re} \Pi_l \nonumber \\
\label{sp}
\end{eqnarray}
and where $f=(\exp \beta(\omega_p-\mu) +1 )^{-1}$ and $n(\omega_q)=(\exp\beta
\omega_q -1)^{-1}$ are Fermi and Bose occupation factors.  The sum over
internal degrees of freedom includes in $S_f$ a sum over states with helicity
equal ($+$) and opposite to ($-$) their chirality, while it involves in $S_b$
a sum over the two transverse and the longitudinal polarizations.  We note
that in the expressions above only the temperature derivatives acting on
occupation factors need to be retained.  All other temperature dependences
cancel out as a consequence of the stationarity of $\Omega$.~\cite{VanBaym}
The terms $S_f$ and $S_b$ represent the contributions from the electron and
photon elementary modes. The term $S'$, which describes residual interactions
between quasiparticles, has the particular property that it vanishes at
one-loop order,~\cite{VanBaym} corresponding to the $\Phi$-diagram and
the self-energies in Fig.~1.  In the following, we will stay at this
level of approximation and neglect $S'$. An example of a higher-order
interaction term $S'$ appears in the description of the thermodynamic
properties of $^3$He, where it describes the effects of
repeated scattering of particle-hole pairs and gives rise to the $T^3 \log
T$ term in the low temperature specific heat.~\cite{He3}

\vspace{-.2cm}
\begin{figure}[ht]
\centerline{
\epsfysize=3cm
\epsfbox{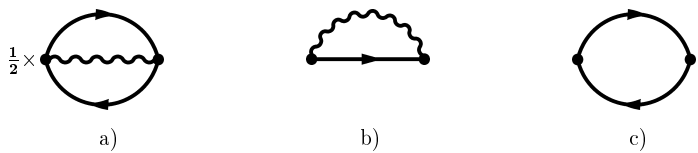}}
\caption{(a) $\Phi$, (b) $\Sigma$ and (c) $\Pi$, to one-loop order.}
\end{figure}

\section{QUASIPARTICLE ANALYSIS OF THE ENTROPY OF AN ELECTROMAGNETIC
PLASMA}

\subsection{A quasiparticle description of the thermodynamics of a QED plasma}

    The reasons for choosing the entropy as the thermodynamic quantity to be
analyzed in terms of its quasiparticle content are twofold.  First, a simple
inspection of Eqs.~(\ref{sf})-(\ref{sp}) shows that, because of the presence
of occupation factors in the integrand, the entropy does not suffer from
ultraviolet divergences.  Potential renormalization problems are thus avoided
as all other thermodynamic quantities can be obtained from the entropy by
simple integration.  Second, a spectral representation can be introduced by
recognizing that
\begin{eqnarray}
{\partial f\over \partial T}&=&-{\partial \sigma_f\over\partial
\omega_p} (\omega_p), \label{dft:sigmaf}\quad {\partial n\over \partial
T}=-{\partial \sigma_n\over\partial \omega_q} (\omega_q),
\label{dnt:sigman}
\end{eqnarray}
where $\sigma_f\equiv -f \log f-(1-f)\log(1-f)$ and $\sigma_n\equiv -n\log n
+(1+n)\log(1+n)$ are the entropy contributions from a single electron mode with
energy $\omega_p$ and from a single photon mode with frequency $\omega_q$.
Integrating Eqs.~(\ref{sf}) and~(\ref{sb}) by parts, we obtain the spectral
representations
\begin{eqnarray}
S_f&=&\sum_{p}\int {d\omega_p\over 2\pi}\,\sigma_f(\omega_p)\,
A_s(\omega_p,p),
\,\,
S_b=\sum_{q}\pp_0^\infty {d\omega_q\over
2\pi}\,\sigma_b(\omega_q) B_s(\omega_q,q)
\label{sbspec}
\end{eqnarray}
where the spectral density $A_s$ is defined as
\begin{eqnarray}
A_s(\omega_p,p)&=&{\sum_\pm\,}{\partial\over\partial \omega_p}\left\{{\rm Re}
G_\pm\,{\rm Im} \Sigma_\pm + 2 {\rm Im}\,\log (-G_\pm)\right\},
\label{as}
\end{eqnarray}
and $B_s$ has a similar structure in terms of $D_l$ and $\Pi_l$.

The physical content of the spectral densities can be elucidated as follows.
For brevity, we only consider $A_s$. If the system develops well-defined
excitations, the fermion propagator takes the following form near a
quasiparticle pole:
\begin{eqnarray}
G_\pm(\omega_p+i 0^+,p) \sim \frac{Z_\pm}{\omega_p-\varepsilon^\pm_p+i
  \Gamma^\pm_p/2 },
\end{eqnarray}
where $Z_\pm$ is the residue of the quasiparticle with energy
$\varepsilon_\pm$ and inverse lifetime $\Gamma_p^\pm=-2 {\rm Im} \Sigma_\pm$.
When the variation of $\Gamma_\pm$ around the quasiparticle pole is
negligible, the density $A_s$ near the pole takes the form
\begin{eqnarray}
A_{s\pm}(\omega_p,p)&\simeq& \frac{(Z_{\pm} \Gamma_{\pm})^3/2}
{((\omega_p-\varepsilon_p^{\pm})^2+(Z_{\pm} \Gamma_{\pm}/2)^2)^2}.
\label{aspole}
\end{eqnarray}
Thus, in the non-interacting limit $\Gamma_\pm \to 0$, $A_s$ reduces to a
delta function and $S_f$ in Eq.~(\ref{sbspec}) coincides with the entropy of a
gas of free electrons with energies $\varepsilon_p^\pm$.  When $\Gamma_\pm$ is
different from zero but small, so that well-defined quasiparticle modes exist,
$A_s$ is a sharply peaked function which falls off faster than a Lorentzian.
The excited quasiparticles as well as the degrees of freedom associated with
their finite lifetimes thus contribute to the entropy $S_f$.  The form of
$B_s$ close to photon quasiparticles is qualitatively similar to that of
$A_s$.  Both $A_s$ and $B_s$ also have support over wide bands of continuum
states, which describe the contributions from Landau damping effects.  In
higher order approximations, the continuum bands also include multiparticle
states.

\subsection{Comparison to perturbative expansions}

The present $\Phi$ conserving approximations also provide a general
framework for organizing calculations in perturbation theory.  To
illustrate we comment on the effects of the
infrared structure and the presence of a gap in the fermion spectrum
on the plasma thermodynamics.  As mentioned earlier, the lack of
static screening of magnetic fields is responsible for a logarithmic
divergence of the damping rate evaluated in perturbation theory,
$\Gamma_\pm(\omega_p) \sim g^2 T \log (g T /
|\omega_p-\varepsilon^\pm_p|)$.  This behavior implies
that $A_s$ {\it vanishes} at the quasiparticle pole, since $A_s \sim
1/\Gamma \sim 1/\log (g T/ |\omega_p-\varepsilon_p^\pm|)$ as $\omega_p
\sim \varepsilon_p^\pm$. There is thus no well-defined quasiparticles
at this level of approximation; a correct quasiparticle description of
the plasma thermodynamics must include a proper treatment of the
long-ranged gauge fields for frequencies close to the poles
$\varepsilon^\pm_p$.

    The effects of the gap in the fermion spectrum on thermodynamic quantities
can also readily be estimated.  Inspection of $S_f$ in Eq.~(\ref{sf}) shows
that low momentum modes with $\omega_p \sim p \sim g T$ contribute a factor
$g^5 T^3$ to the entropy per unit volume, in contrast with the contribution
from the long-wavelength plasmon modes, $S_{\rm plasmons}\sim g^3
T^3$.~\cite{plasmons} The difference in the order of magnitude can be traced
back to statistics.  For low frequencies $\omega_p, \omega_q \sim g T$, the
temperature derivative $\partial f/\partial T\sim \omega_p/T^2$ in
Eq.~(\ref{sf}) is two powers of g smaller than the derivative $\partial
n/\partial T \sim 1/\omega_q$ in Eq.~(\ref{sb}).  Hence $S_f \sim g^2 S_{\rm
plasmons}$.

    Further comparisons between our expressions and the standard results of
perturbation theory reveal relations that follow from the self-consistent
character of the derivation.  For the formalism to reproduce all terms in the
plasmon contribution to the entropy, $\sim g^3 T^3$, one needs to include
contributions from both long wavelength photons, via $S_b$, and electron
modes, via $S_f$.  These last terms describe the effects of the plasmons on
the single particle energy of the electrons, via the real part of the
self-energy, and are automatically included in a self-consistent treatment.
The approach presented in this talk has recently been applied to purely
gluonic QCD by Blaizot et al.,~\cite{SQCD} who have succeeded in obtaining an
accurate quasiparticle description of lattice data for the equation of state
down to temperatures twice $T_c$.  There again, it is important to take into
account the effects of soft modes on the dispersion relation of the hard modes
to reproduce the entropy correctly up the order $g^3$.  A description of these
effects in terms of microscopic processes can be found in Ref.~\cite{SQCD}.

\section{CONCLUSIONS}

    In this talk, we generalized the concept of conserving approximations to
relativistic theories.  We have illustrated the technique for a hot
electromagnetic plasma and shown how the approach allows one to resolve the
entropy into contributions from the elementary degrees of freedom.  We have
also pointed out the advantages of implementing $\Phi$-conserving techniques
as a basis for perturbation expansions.

\section*{Acknowledgements}
    This research was supported in part by NSF Grants PHY98-00978 and
PHY94-21309.

\end{document}